\documentclass[prl,twocolumn,superscriptaddress]{revtex4}

\usepackage{epsfig}

\begin{document}

\title{Possible experimental manifestations of the many-body
  localization}
\author{D.~M.~Basko}
\email{basko@phys.columbia.edu}
\affiliation{Physics Department, Columbia University, New York, NY
10027, USA}
\author{I.~L.~Aleiner}
\affiliation{Physics Department, Columbia University, New York, NY
10027, USA}
\author{B.~L.~Altshuler}
\affiliation{Physics Department, Columbia University, New York, NY
10027, USA}
\affiliation{NEC-Laboratories America,
Inc., 4~Independence Way, Princeton, NJ 085540 USA}

\begin{abstract}
Recently, it was predicted that if all one-electron states in a
non-interacting disordered system are localized, the interaction
between electrons in the absence of coupling to phonons leads to a
finite-temperature metal-insulator transition. Here, we show that
even in the presence of a weak coupling to phonons the transition
manifests itself (i)~in the nonlinear conduction, leading to a
bistable $I$-$V$ curve, (ii)~by a dramatic enhancement of the
nonequilibrium current noise near the transition.
\end{abstract}

{\maketitle}

{\em Introduction.---}
Low-temperature charge transport in disordered conductors is
governed by the interplay between elastic scattering of electrons off
static disorder (impurities) and inelastic scattering
(electron-electron, electron-phonon, etc.). At low dimensions an
arbitrarily weak
disorder localizes~\cite{Anderson58} all
single-electron states~\cite{Gertsenshtein,Abrahams}, and
there would be no transport without inelastic processes. For the
electron-phonon scattering, the 
dc conductivity $\sigma(T)$ at low temperatures~$T$
is known since long ago~\cite{Mott}: in
$d$ dimensions
\begin{equation}\label{Mott=}
\ln\sigma(T)\propto-1/T^\gamma,\quad\gamma=1/(d+1)\,.
\end{equation}
What happens if the
only possible inelastic process is electron-electron scattering?
The answer to this question was found only recently~\cite{us}:
$\sigma(T)=0$ identically for $T<T_c$, the temperature of a
metal-insulator transition. Here we discuss experimental
manifestations of this transition in real systems, where both
electron-electron and electron-phonon interactions are present.
We show that (i)~the $I$-$V$ characteristic exhibits
a bistable region, and (ii)~non-equilibrium current noise is
enhanced near~$T_c$.

The notion of localization was originally introduced for a single
quantum particle in a random
potential~\cite{Anderson58}. Subsequently, the concept of Anderson
localization was shown to manifest itself in a broad variety of
phenomena in quantum physics.
%
This concept can also be extended to many-particle
systems. Statistical physics of many-body systems is based on the
microcanonical distribution, i.~e., all states with a given energy are
assumed to be realized with equal probabilities.
This assumption means {\em delocalization} in the space of possible
states of the system. It does not hold for non-interacting particles;
however, it is commonly believed that  an arbitrarily weak interaction
between the particles eventually equilibrates the system and
establishes the microcanonical distribution.

Many-body dynamics of interacting systems and its relation to Anderson
localization has been discussed in the context of
nuclear~\cite{Agassi} and molecular~\cite{Wolynes} physics. For
interacting electrons in a chaotic quantum dot this issue was raised
in Ref.~\cite{AGKL}, where it was shown that electron-electron
interaction may not be able to equilibrate the system. This
corresponds to Anderson
{\em localization} in the many-body space. Recently it was
demonstrated that in an infinite low-dimensional system of (weakly)
interacting electrons, subject to a static disorder, Anderson
transition in the many-body space manifests itself as a
finite-temperature metal-insulator transition~\cite{us}.

\begin{figure}
\includegraphics[width=8cm]{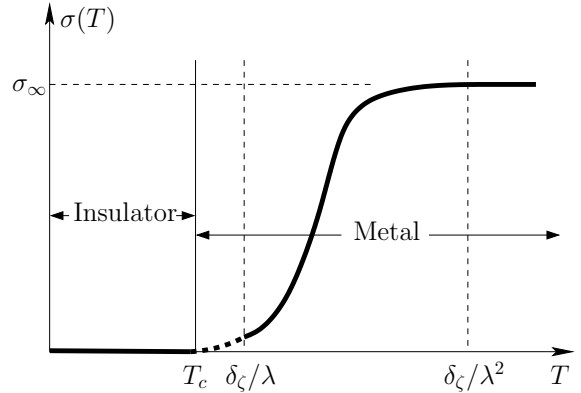}
\caption{Schematic temperature dependence of the dc conductivity
  $\sigma(T)$ for electrons subject to a disorder potential localizing
  all single particle eigenstates, in the presence of weak short-range
  electron-electron $\lambda\delta_\zeta$, $\lambda\ll{1}$,
  established in Ref.~\cite{us}. Below the point of the
  many-body metal-insulator transition,
  $T<T_c\sim\delta_\zeta/|\lambda\ln\lambda|$, no inelastic relaxation
  occurs and $\sigma(T)=0$. At $T\gg\delta_\zeta/\lambda^2$ the
  high-temperature metallic perturbation theory\cite{AA} is valid, and
  corrections to the Drude conductivity~$\sigma_\infty$ are small. In
  the interval $\delta_\zeta/\lambda\ll{T}\ll\delta_\zeta/\lambda^2$
  electron-electron interaction leads to electron transitions between
  localized states, and the conductivity depends on temperature as a
  power-law.
}\label{fig:summary}
\end{figure}

Let single-particle eigenstates be localized
on a spatial scale~$\zeta_{loc}$ (localization length). The
characteristic energy scale of the problem is the level spacing within
the localization volume: $\delta_\zeta=1/(\nu\zeta_{loc}^d)$,
$\nu$~being the density of states per unit volume. Below we neglect
energy dependences of $\nu$, $\zeta_{loc}$, and $\delta_\zeta$.
According to Ref.~\cite{us}, as long as electrons are not coupled
to any external bath (such as phonons), a weak short-range
electron-electron interaction (of typical magnitude
$\lambda\delta_\zeta$, with the dimensionless coupling constant
$\lambda\lesssim{1}$) does not cause inelastic relaxation
unless the temperature~$T$ exceeds a critical value:
\begin{equation}\label{Tc=}
T_c\sim\frac{\delta_\zeta}{|\lambda\ln\lambda|}.
\end{equation}
The small denominator $|\lambda\ln\lambda|$ represents the
characteristic matrix element of the creation of an electron-hole
pair. The ratio $T_c/\delta_\zeta$ is the number of states available
for such a pair (in other words, the phase volume) at $T=T_c$. Only
provided that this number is large enough to compensate the smallness
of the matrix element, the interaction delocalizes the many-body states
and thus leads to an irreversible dynamics.
As a result, the finite-temperature dc conductivity $\sigma(T)$
vanishes identically if $T<T_c$, while $\sigma(T>T_c)$ is finite,
i.~e. at $T=T_c$ a metal-insulator transition occurs. The
overall dependence $\sigma(T)$ is summarized in
Fig.~\ref{fig:summary}.

\begin{figure}
\includegraphics[width=8cm]{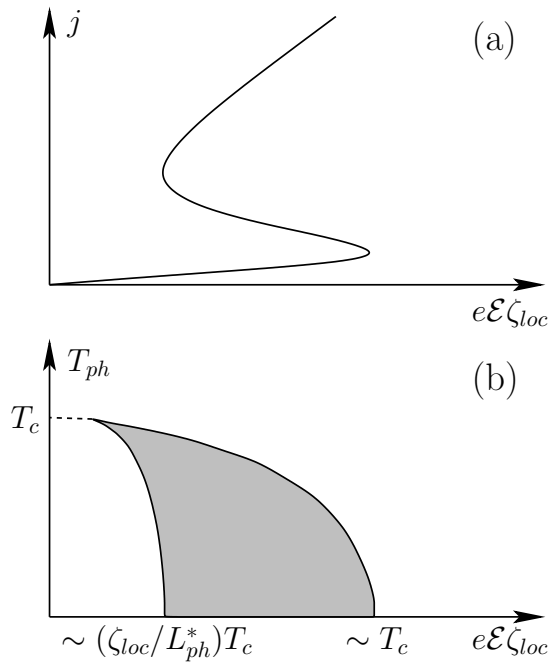}
\caption{(a)~sketch of the bistable $I$-$V$ curve for a fixed
value of~$T_{ph}$; (b)~$(\mathcal{E},T_{\rm ph})$-plane with the
bistable region schematically shown by shading. The dashed line
represents the crossover between the metallic state at high
electric field~$\mathcal{E}$ or high phonon temperature~$T_{ph}$,
and the insulating state at low~$\mathcal{E}$ and low~$T_{ph}$.}
\label{fig:bistable}
\end{figure}

In any real system the electron-phonon interaction is always
finite. This makes $\sigma(T)$ finite even at $T<T_c$: $\sigma(T)$ is
either exponentially small [Eq.~(\ref{Mott=})] at $T\ll\delta_\zeta$,
or follows a power-law~\cite{GMR} at $\delta_\zeta\ll{T}\ll{T}_c$. At
the transition point the phonon-induced conductivity is not
exponentially small, i.~e. phonons smear the transition into a
crossover. Are there any experimental signatures of the many-body
localization? In what follows we show that if the electron-phonon
coupling is weak enough, a {\em qualitative} signature of the
metal-insulator transition can be identified in the nonlinear
conduction. Namely, in a certain interval of applied electric
fields~$\mathcal{E}$ and phonon temperatures~$T_{ph}$ both metallic
and insulating states of the system turn out to be stable! As a
result, the $I$-$V$ curve exhibits an S-shaped bistable region
(Fig.~\ref{fig:bistable}). Moreover, we show that the many-body
character of the electron conduction dramatically modifies the
non-equilibrium noise near the transition
[Eq.~(\ref{Fanocritical=})].

{\em Bistable $I$-$V$ curve.---}
Our arguments are based on two observations.
First, in the absence of phonons a weak but finite electric field
cannot destroy the insulating state -- it rather shifts the transition
temperature. Let us neglect the effect of the field on the
single-particle wave functions, representing it as a tilt of the local
chemical potential of electrons. Then at $T=0$ the role of the field
in the insulating regime is increase the energy of the electron-hole
(e-h) excitation of a size~$L$ by $e\mathcal{E}L$. This
provides in additional phase volume of the order of
$e\mathcal{E}L/\delta_\zeta$. However, for $L>\zeta_{loc}$ the matrix
element for creation of such an excitation quickly vanishes. In the
diagrammatic language for the effective model of Ref.~\cite{us} this
means that {\em each} electron-electron interaction vertex must be
accompanied by tunneling vertices which describe coupling between
localization volumes and whose number is (i)~at least one in order to
gain phase volume (in contrast to the
finite-$T$ case when tunneling had to be included only to overcome
the finiteness of the phase space in a single grain~\cite{us}), and
(ii)~not much greater than one, otherwise the diagram is exponentially
small. As a result, at $T=0$  the insulator state is stable provided
that $\mathcal{E}<\mathcal{E}_c\sim{T}_c/(e\zeta_{loc})$.

In the same way one can analyze the finite-temperature correction to
the critical field, and the finite-field correction to the critical
temperature can be found by taking into account the extra phase
volume in the calculation of Ref.~\cite{us}. One obtains $\sigma(T)=0$
for $T<T_c(\mathcal{E})$, where
\begin{equation}\label{Tcshift=}
T_c(\mathcal{E})=T_c-c_1{e}\mathcal{E}\zeta_{loc}\,.
\end{equation}
with a model-dependent factor $c_1\sim{1}$, weakly dependent
on~$\mathcal{E}$ [here and below $T_c$ without the
argument~$\mathcal{E}$ is the
zero-field value given by Eq.~(\ref{Tc=})].
As a consequence, at $T<{T}_c(\mathcal{E})$ the nonlinear transport,
as well as the linear one, has to be phonon-assisted.

The second observation is that when both $\sigma$~and~$\mathcal{E}$
are finite, there is Joule heating. The thermal balance is
qualitatively different in the insulating and the metallic
phases. Deep in the insulating phase ($T\ll{T}_c$) each electron
transition is accompanied by a
phonon emission/absorption, i.~e. electrons are always in
equilibrium with phonons whose temperature~$T_{ph}\ll{T}_c$ we
assume to be fixed.
On the contrary, in the metallic phase electrons gain
energy when drifting in the electric
field, i.~e. they are heated. Due to this Joule heating the effective
electron temperature~$T_{el}$ deviates from the bath
temperature. The role of phonons is then to
stabilize~$T_{el}$. For weak electron-phonon coupling $T_{el}$ and
$T_{ph}$ can differ significantly. 
A self-consistent estimate for~$T_{el}$ follows from
\begin{eqnarray}\label{Tel=}
&&T_{el}-T_{ph}\sim{e}\mathcal{E}L_{ph}(T_{el})\,,\\
&&L_{ph}(T_{el})=\sqrt{D(T_{el})\,\tau_{ph}(T_{el})}\,.
\end{eqnarray}
Here $\tau_{ph}(T_{el})$ is the time it takes an electron to emit or
absorb a phonon, $L_{ph}(T_{el})$ is the typical electron displacement
during this time, and $D(T_{el})=\sigma(T_{el})/(e^2\nu)$ is the
electron diffusion coefficient.

\begin{figure}
\includegraphics[width=8cm]{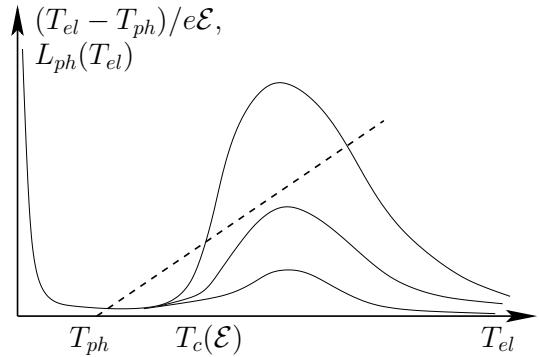}
\caption{\label{fig:plot}
Sketch of the dependences $(T_{el}-T_{ph})/({e}\mathcal{E})$ (dashed
line) and $L_{ph}(T_{el})$ (solid lines, electron-phonon coupling
strength being weaker for higher curves). Actual value of~$T_{el}$ is
determined by their crossing.}
\end{figure}

We sketch in Fig.~\ref{fig:plot} $(T_{el}-T_{ph})/({e}\mathcal{E})$
and $L_{ph}(T_{el})$ for different electron-phonon coupling strengths
as functions of~$T_{el}$. It is taken into account that
(i)~$L_{ph}$ coincides with variable range hopping length at
$T_{el}\ll\delta_\zeta$,
(ii)~$L_{ph}\sim\zeta_{loc}$~at
$\delta_\zeta\ll{T}_{el}\ll{T}_c(\mathcal{E})$,
(iii)~$D(T_{el})$ quickly rises to its large metallic value
$D_\infty\sim\delta_\zeta\zeta_{loc}^2$ near $T_c(\mathcal{E})$,
(iv)~$\tau_{ph}(T_{el})$ decreases as a power law with
increasing~$T_{el}$.
The peak of the curve rises with decreasing electron-phonon coupling
strength, and eventually the curve crosses the straight line. After
that, in addition to $T_{el}=T_{ph}$, Eq.~(\ref{Tel=}) acquires two
more solutions, both with $T_{el}>T_{c}(\mathcal{E})$, of which only
the rightmost solution is stable. The maximum of $L_{ph}$ can be
estimated as $L_{ph}^*\sim\sqrt{D_\infty\tau_{ph}(T_c)}$, so three
solutions appear when
${T}_c(\mathcal{E})-T_{ph}\ll{e}\mathcal{E}L_{ph}^*$.
At the same time, as seen from Eq.~(\ref{Tcshift=}), electric
field is unable to break down the insulator as long as
${e}\mathcal{E}\zeta_{loc}\ll{T}_c-T_{ph}$. Thus, the
interval of electric fields where both regimes are stable, is
determined by
\begin{equation}
\frac{{T}_c(\mathcal{E})-T_{ph}}{eL_{ph}^*}\ll\mathcal{E}
\ll\frac{{T}_c-T_{ph}}{e\zeta_{loc}}\,.
\label{bistable=}
\end{equation}
The two conditions are compatible provided that
\begin{equation}
L_{ph}^*\gg\zeta_{loc}\,,
\end{equation}
which is realistic when electron-phonon coupling is weak.

In the bistable region~(\ref{bistable=}), for a given value
of~$\mathcal{E}$ one finds two stable solutions for~$T_{el}$, giving
two possible values of the conductivity and the current, which
corresponds to an $S$-shape current-voltage
characteristic~\cite{Scholl}, the third (unstable) solution
corresponding to the negative differential conductivity branch. The
macroscopic consequences of such behavior depend on the
dimensionality. In a 2d sample the two phases of different electronic
temperature and current density can coexist, separated by a boundary
of the width $\sim{L}_{ph}^*$, parallel to the direction of the
electric field.
The particular state of the system determined by the boundary
conditions (properties of the external circuit), as well as by the
history.

{\em Noise enhancement.---} In the vicinity of the critical point
conduction is dominated by correlated many-electron transitions
(electronic cascades). Each cascade is triggered by a single
phonon. As $T_{el}\to{T}_c(\mathcal{E})$, the typical value~$\bar{n}$
of the number~$n$ of electrons in the cascade diverges together with
the time duration of a cascade.
The results of Ref.~\cite{us}, adapted for a finite electric field,
give the following probability for an $n$-electron transition
to go with the rate~$\Gamma$:
\begin{equation}\label{rates=}
P_n(\Gamma)=\sqrt{\frac{\bar\Gamma_n}{4\pi}}
\frac{e^{-\bar\Gamma_n/(4\Gamma)}}{\Gamma^{3/2}}\,,\quad
\bar\Gamma_n\propto
\left(\frac{T_{el}+c_1e\mathcal{E}\zeta_{loc}}{T_c}\right)^{2n},
\end{equation}
which gives
\begin{equation}\label{nbar=}
1/\bar{n}\sim\ln[T_c/(T_{el}+c_1e\mathcal{E}\zeta_{loc})].
\end{equation}
The divergence in~$\bar{n}$ is cut off when
electron-phonon coupling is finite. The largest~$\bar{n}$ is
such that the phonon broadening of the single-electron levels,
$1/\tau_{ph}(T_c)$, is comparable to $\bar{n}$-particle level
spacing (in other words, time duration of a cascade cannot exceed
$\tau_{ph}$):
\begin{equation}\label{phononcutoff=}
\frac{1}{\tau_{ph}(T_c)}\sim\delta_\zeta
\left(\frac{\delta_\zeta/\bar{n}^{\alpha{d}}}{{T}_c}\right)^{4\bar{n}}
\;\;\;\Rightarrow\;\;\;
\bar{n}_{max}
\sim\frac{1}{4}
\frac{\ln(\delta_\zeta\tau_{ph})}{\ln(T_c/\delta_\zeta)}\,,
\end{equation}
with logarithmic precision; $\bar{n}^\alpha$ represents the
divergent spatial extent of the cascade (correlation
length)~\cite{index}.

Each many-electron transition can be characterized, besides its
rate~$\Gamma$, by the total dipole moment~$\vec{d}$ it produces.
The corresponding backward transition produces the dipole moment
$-\vec{d}$ and goes with the rate
$\Gamma{e}^{-(\vec{\mathcal{E}}\cdot\vec{d})/T_{el}}$~\cite{localfield}.
The average current $\langle{I}(t)\rangle$ is determined by the
difference between forward and backward rates; obviously, it
vanishes for $\mathcal{E}=0$. At the same time, the noise power
(second cumulant)
$S_2\equiv\int[\langle{I(t)I(t')}\rangle
-\langle{I(t)}\rangle\langle{I(t')}\rangle]\,dt'$ is determined by the
sum of the forward and backward rates; at $\mathcal{E}=0$ it is
given by the equilibrium Nyquist-Johnson expression.



Equilibrium noise carries no information about the nature of
conduction. To see a signature of many-electron transitions it
would be natural to analyze the shot noise, whose power is
proportional to the charge transferred in a single event.
Many-electron cascades would then correspond to ``bunching'' of
electrons, thus increasing the shot noise. However, shot noise is
observed in the limit when transitions transferring charge only in
one direction (namely, $\vec{d}\cdot\vec{\mathcal{E}}>0$) are
allowed, i.~e. $T_{el}\ll{e}\mathcal{E}\zeta_{loc}$, which
is impossible to satisfy in the insulating state,
as $T_{el}\sim\max\{T_{ph},e\mathcal{E}\zeta_{loc}\}$~\cite{Levin}.
Thus, $S_2$~inevitably has both equilibrium and non-equilibrium
contributions, which are difficult to separate.

To see the ``bunching'' effect unmasked by a large thermal noise at
low fields one should study the third Fano factor~$S_3$ of the current
fluctuations~\cite{Levitov}. Indeed, being proportional to an odd
power of the current, it vanishes in equilibrium, so it is not subject
to the problems described above for~$S_2$. In a wire of length~$L$ the
ratio $S_3/\langle{I}\rangle$ is given by
\begin{equation}\label{Fano=}
\frac{S_3}{\langle{I}\rangle}=
\frac{L^{-3}\left\langle\!\left\langle
\Gamma{d}^3\right\rangle\!\right\rangle}
{L^{-1}\left\langle\!\left\langle
\Gamma{d}\right\rangle\!\right\rangle}\,.
\end{equation}
The double angular brackets on the right-hand side
mean the sum over all allowed transitions. For nearest-neighbor
single-electron transitions with $d=\pm{e}\zeta_{loc}$
Eq.~(\ref{Fano=}) gives
$S_3/\langle{I}\rangle\sim{e}^2(\zeta_{loc}/L)^2$, which is analogous
to the Schottky expression reduced by the effective number of tunnel
junctions in series, $L/\zeta_{loc}$, for~$S_2$~\cite{Landauer}.

Since $d^3$ diverges stronger than~$d$ as $\bar{n}\to\infty$, we
expect a divergence in Eq.~(\ref{Fano=}). The critical index of~$d$
depends on the order of limits: $d\sim\sqrt{n}{e}\zeta_{loc}$ if the
linear response limit $\mathcal{E}\to{0}$ is taken prior to
$\bar{n}\to\infty$, while
$d\sim{n}{e}\zeta_{loc}(e\mathcal{E}\zeta_{loc}/T_c)$ for a small but
finite~$\mathcal{E}$. As a result,
\begin{equation}\label{Fanocritical=}
\frac{S_3}{\langle{I}\rangle}\sim\left(\frac{e\zeta_{loc}}L\right)^2
\max\left\{\bar{n}\,,
\left(\frac{\bar{n}e\mathcal{E}\zeta_{loc}}{T_c}\right)^2\right\},\quad
\bar{n}\lesssim\bar{n}_{max}\,,
\end{equation}
where $\bar{n}$ is given by Eq.~(\ref{nbar=}), and
the saturation of the divergence is determined by
Eq.~(\ref{phononcutoff=}). Upon further increase of the temperature,
the system crosses over to the metallic state, and $\bar{n}$~starts
to decrease. This decrease is governed by the same
Eq.~(\ref{phononcutoff=}) with the phonon inelastic rate substituted
by the typical value of the electron-electron inelastic rate, which
grows with temperature. As the critical behavior of~$\Gamma$ on the
metallic side of the transition is unknown, we cannot give any
quantitative estimate of~$S_3$ above~$T_c$.

{\em Conclusions.---} In conclusion, we have shown that the
finite-temperature metal-insulator transition, predicted
theoretically in Ref.~\cite{us}, can manifest itself on the
macroscopic level as an S-shape current-voltage characteristic with a
bistable region.
In fact, the hysteretic behaviour of the current in
Y$_x$Si$_{1-x}$~\cite{Ladieu} is a possible candidate for the effect
discussed in the present paper.

Besides, we have shown that the many-body nature of the conduction
near the transition manifests itself in the dramatic increase of the
non-equilibrium current noise: the noise depends on the total charge
transferred in each random event, while the number of  electrons,
involved in such an event, increases as one approaches the
transition.

We acknowledge discussions with M.~E.~Gershenson,
C.~M.~Marcus, A.~K.~Savchenko, M.~Sanquer, and thank H.~Bouchiat for
drawing our attention to Ref.~\cite{Ladieu}.


\begin{thebibliography}{99}
\bibitem{Anderson58}
P.~W.~Anderson, Phys. Rev. {\bf 109}, 1492 (1958).
\bibitem{Gertsenshtein}
M.~E.~Gertsenshtein and V.~B.~Vasil'ev, Teoriya Veroyatnostei i ee
Primeneniya {\bf 4}, 424 (1958) [Theory of Probability and its
Applications {\bf 4}, 391 (1958)].
\bibitem{Abrahams}
E.~Abrahams, P.~W.~Anderson, D.~C.~Licciardello, and
T.~V.~Ramakrishnan, Phys. Rev. Lett. {\bf 42}, 673 (1979).
\bibitem{Mott}
N.~F.~Mott, J. Non-Cryst. Solids {\bf 1}, 1 (1968); Mott formula does
not into account Coulomb interaction, which changes the value of
$\gamma$ in Eq.~(\ref{Mott=}) [A.~L.~Efros and B.~I.~Shklovskii,
  J.~Phys.~C {\bf 8}, L49 (1975)].
\bibitem{us}
D.~M.~Basko, I.~L.~Aleiner, and B.~L.~Altshuler, Ann. Phys. {\bf
321}, 1126 (2006). See also cond-mat/0602510.
\bibitem{Agassi}
D.~Agassi, H.~A.~Weidenm\"uller, and G.~Mantzouranis, Phys. Rep.
{\bf 22}, 145 (1975).
\bibitem{Wolynes}
D.~E.~Logan and P.~G.~Wolynes, J.~Chem. Phys. {\bf  93}, 4994
(1990).
\bibitem{AGKL}
B.~L.~Altshuler, Y.~Gefen, A.~Kamenev, and L.~S.~Levitov, Phys.
Rev. Lett. {\bf 78}, 2803 (1997).
\bibitem{AA}
B.~L.~Alshuler and A.~G.~Aronov, in {\it Electron-Electron
Interactions in Disordered Systems}, ed. by A.~L.~Efros and
M.~Pollak (Elsevier, Amsterdam, 1985).
\bibitem{GMR}
A.~A.~Gogolin, V.~I.~Melnikov, and \'{E}.~I.~Rashba, Zh. Eksp.
Teor. Fiz. {\bf 69}, 327 (1975) [Sov. Phys. JETP {\bf 42}, 168
(1976)].
\bibitem{Scholl}
E. Sch\"oll, {\it Nonlinear Spatio-Temporal Dynamics and Chaos in
  Semiconductors}
(Cambridge University Press, Cambridge, 2001).
\bibitem{index}
The index~$\alpha$ depends on the dimensionality only. Arguments
given in Sec.~6.5 of Ref.~\cite{us} fix $\alpha=1$ in 1d, and restrict
$1/2\leq\alpha\leq{3}/4$ in 2d, $1/3\leq\alpha<3/5$ in 3d,
$1/d\leq\alpha\leq{1}/2$ in higher dimensions.
\bibitem{localfield}
Strictly speaking, the electric field appearing in
Eq.~(\ref{rates=}) is the {\em local} electric field, resulting
from the spatial redistribution of charge over the random resistor
network. Here we neglect fluctuations of the local field (more
precisely, their correlations with the spatial fluctuations in
$\Gamma$~and~$\vec{d}$), and understand ~$\mathcal{E}$ as the
external field. This is a good approximation for the
distribution~(\ref{rates=}),
as the current pattern is determined by the typical resitors,
$\Gamma\sim\bar\Gamma_n$.
\bibitem{Levin}
B.~I.~Shklovskii, Fiz. Tekh. Poluprovodn. {\bf 6}, 2335 (1972)
[Sov. Phys. Semicond. 6, 1964 (1973)];
B.~I.~Shklovskii, E.~I.~Levin, H.~Fritzsche, and S.~D.~Baranovskii, in
{\em Transport, Correlation and Structural Defects}, edited by
H. Fritzsche (World Scientific, Singapore, 1990).
\bibitem{Levitov}
L.~S.~Levitov and M.~Reznikov, cond-mat/0111057; Phys. Rev.~B
{\bf 70}, 115305 (2004).
\bibitem{Landauer}
R.~Landauer, Physica~B {\bf 227}, 156 (1996); A.~N.~Korotkov and
K.~K.~Likharev, Phys. Rev.~B {\bf 61}, 15975 (2000).
\bibitem{Ladieu}
F.~Ladieu, M.~Sanquer, and J.~P.~Bouchaud, Phys. Rev.~B {\bf 53}, 973
(1996).
\end{thebibliography}
\end{document}